\documentclass{IEEEtran4PSCC}

\usepackage{cite}

\ifCLASSINFOpdf
   \usepackage[pdftex]{graphicx}
  \DeclareGraphicsExtensions{.pdf,.jpeg,.png}
\else
  \usepackage[dvips]{graphicx}
  \DeclareGraphicsExtensions{.eps}
\fi
\usepackage[cmex10]{amsmath}
\usepackage[inline,shortlabels]{enumitem}
\usepackage{cleveref}
\usepackage{amsthm}
\usepackage{siunitx}
\usepackage{hhline}
\usepackage{multirow}
\usepackage{booktabs}
\usepackage{subcaption}
\usepackage{threeparttable}
\usepackage{etoolbox}

\crefname{equation}{}{}
\Crefname{equation}{Equation}{Equations}
\crefrangelabelformat{equation}{(#1)-(#2)}
\crefname{section}{§\!\!}{§§\!\!}
\Crefname{section}{Section}{Sections}
\crefname{figure}{Fig.}{Figs.}      
\Crefname{figure}{Figure}{Figures}
\crefname{table}{Table}{Tables}      
\Crefname{table}{Table}{Tables}

\newtoggle{TRACKCHANGES}

\iftoggle{TRACKCHANGES}{

\newcommand{\myadd}[1]{\textcolor{blue}{#1}}
}{

\newcommand{\myadd}[1]{#1}
}

\makeatletter
\let\old@ps@headings\ps@headings
\let\old@ps@IEEEtitlepagestyle\ps@IEEEtitlepagestyle
\def\psccfooter#1{%
    \def\ps@headings{%
        \old@ps@headings%
        \def\@oddfoot{\strut\hfill#1\hfill\strut}%
        \def\@evenfoot{\strut\hfill#1\hfill\strut}%
    }%
    \def\ps@IEEEtitlepagestyle{%
        \old@ps@IEEEtitlepagestyle%
        \def\@oddfoot{\strut\hfill#1\hfill\strut}%
        \def\@evenfoot{\strut\hfill#1\hfill\strut}%
    }%
    \ps@headings%
}
\makeatother

\psccfooter{%
        \parbox{\textwidth}{\hrulefill \\ \small{24th Power Systems Computation Conference} \hfill \begin{minipage}{0.2\textwidth}\centering \vspace*{4pt} \includegraphics[scale=0.06]{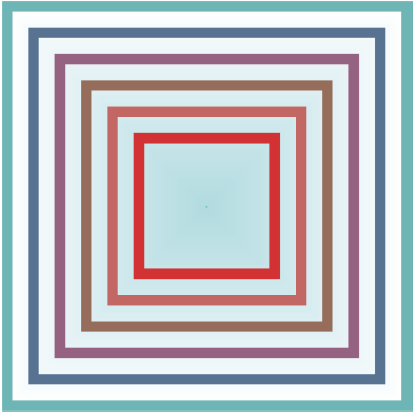}\\\small{PSCC 2026} \end{minipage} \hfill \small{Limassol, Cyprus --- June 8-12, 2026}}%
}

\begin{document}
%
\title{Neural Network-Assisted Model Predictive Control for Implicit Balancing}

\author{\IEEEauthorblockN{Seyed Soroush Karimi Madahi\IEEEauthorrefmark{1},
Kenneth Bruninx\IEEEauthorrefmark{3},
Bert Claessens\IEEEauthorrefmark{1}\IEEEauthorrefmark{2}, 
Chris Develder\IEEEauthorrefmark{1}}
\IEEEauthorblockA{\IEEEauthorrefmark{1} IDLab, Ghent University -- imec, Ghent, Belgium}
\IEEEauthorblockA{\IEEEauthorrefmark{2} Beebop.ai, Belgium}
\IEEEauthorblockA{\IEEEauthorrefmark{3} Faculty of Technology, Policy \& Management, Delft University of Technology, Delft, The Netherlands}
}


\maketitle

\begin{abstract}
In Europe, balance responsible parties can deliberately take out-of-balance positions to support transmission system operators (TSOs) in maintaining grid stability and earn profit, a practice called implicit balancing. Model predictive control (MPC) is widely adopted as an effective approach for implicit balancing. The balancing market model accuracy in MPC is critical to decision quality. Previous studies modeled this market using either (i) a convex market clearing approximation, ignoring proactive manual actions by TSOs and the market sub-quarter-hour dynamics, or (ii) machine learning methods, which cannot be directly integrated into MPC. To address these shortcomings, we propose a data-driven balancing market model integrated into MPC using an input convex neural network to ensure convexity while capturing uncertainties. To keep the core network computationally efficient, we incorporate attention-based input gating mechanisms to remove irrelevant data. Evaluating on Belgian data shows that the proposed model both improves MPC decisions and reduces computational time.
\end{abstract}


\thanksto{\noindent Submitted to the 24th Power Systems Computation Conference (PSCC 2026).}

\section{Introduction}
\label{sec:intro}
The integration of variable renewable energy sources in the power system introduces uncertainty into grid operations, making supply–demand balancing more challenging for transmission system operators (TSOs). In Europe, TSOs maintain grid balance by delegating part of the necessary corrective balancing actions to balance responsible parties (BRPs), such as energy companies. BRPs that fail to balance their portfolio in each imbalance settlement period (ISP) are subject to imbalance prices. Based on the European Electricity Balancing Guidelines (EBGL), all TSOs should apply a single imbalance pricing methodology, in which both BRP shortages and BRP surpluses are exposed to the same price~\cite{ENTSO}. Depending on the system imbalance and the position of the BRP, this results in a net loss or benefit for the BRP. This also implies that BRPs can consciously take out-of-balance positions to support grid stability and gain financial profit, a practice known as implicit or passive balancing. However, implicit balancing is a complex problem due to the high uncertainty and nonlinear nature of the balancing energy market, as well as the partial observability of other BRPs and the TSO's manual actions.

Implicit balancing, leveraging flexible assets such as battery energy storage systems (BESS), has been addressed in the literature through both model-based optimization and model-free reinforcement learning (RL) methods. Studies focusing on optimization methods, primarily model predictive control (MPC), formulate the implicit balancing problem as a bi-level optimization problem: the upper level determines the BRP's profit and the lower level models balancing energy market clearing~\cite{smets2023strategic,wessel2024risk,madahi2025gaming}. On the other hand, RL-based approaches model the implicit balancing problem as a sequential decision-making problem using a Markov decision process: the RL agent is trained through interaction with the environment (in this case, the imbalance market) to take actions at each time step based on the current state, which usually includes real-time and historical information about system imbalance, the grid, and the BRP~\cite{madahi2024distributional,madahi2025risk,ALLY2025126588}. In our previous work~\cite{madahi2025model}, we combined MPC and RL for implicit balancing using a BESS. By leveraging the complementary nature of MPC and RL, the MPC-guided RL method outperformed standalone MPC and RL strategies in terms of profit.

The quality of actions in both optimization-based and MPC-guided RL methods is strongly influenced by the accuracy of the electricity market model. In~\cite{smets2023strategic,wessel2024risk,madahi2025model}, the balancing market used for MPC was modeled based on quarter-hourly balancing energy market clearing. However, this model overlooks the sub-quarter-hourly dynamics of the imbalance settlement mechanism, which play a crucial role in determining imbalance prices~\cite{madahi2025gaming}. Moreover, this model does not account for uncertainties in frequency restoration reserve (FRR) activations as a result of partially known FRR provider availability, grid congestion, and out-of-merit-order manual frequency restoration reserve (mFRR) activations by TSOs, which also influence imbalance prices~\cite{allard2024forecast}. 

Altenatively, imbalance prices can be directly forecasted using machine learning and deep learning~\cite{smets2025value,bottieau2024logic,lucas2020price,plakas2025prediction}, which have been shown to yield more accurate forecasts than fundamental, model-based approaches~\cite{smets2025value}. Smets et al.~\cite{smets2025value} used a value-oriented forecasting method to predict imbalance prices by taking into account the downstream decision problem for which the forecaster is trained. The predicted imbalance prices are directly used by the MPC for decision-making. Yet, this framework ignores the effect of battery actions on imbalance prices, which is particularly significant for large batteries and assets~\cite{bruninx2025day}. To address this issue, the price prediction model must be integrated into the optimization problem. For this purpose, Dolányi et al.~\cite{dolanyi2023capturing} introduced the concept of a mathematical program with a neural network (NN) constraint, in which the lower-level problem (market model) is replaced by a surrogate neural network. State-of-the-art machine learning models to predict real-time and imbalance prices \cite{bottieau2024logic,lucas2020price,plakas2025prediction} are, however, not convex on their inputs, resulting in mixed-integer nonlinear programs (MINLP). These are challenging to solve, which limits their applicability in near-real-time decision making. 

To address these shortcomings, we propose a price-maker NN-assisted MPC framework for implicit balancing. This is achieved by modeling the balancing energy market using an input convex neural network (ICNN). This allows leveraging the powerful capability of NNs to learn nonlinear mappings (such as the influence of sub-quarter-hourly dynamics on imbalance prices), to model market uncertainties (such as proactive mFRR activations), and to capture partially known parameters (such as other market players' behavior), while the ICNN architecture ensures convexity of its output(s) with respect to its input(s). By incorporating the proposed ICNN into the implicit balancing problem, we formulate the optimization problem as a mixed-integer quadratic program (MIQP), in which finding the global optimal solution is guaranteed. 
We furthermore equip the proposed ICNN with attention and embedding layers to further boost its performance. The attention and embedding layers are trained end-to-end with the ICNN to filter FRR bids based on current and historical system imbalances, providing the most relevant bids as input to the ICNN. These trainable preprocessing steps enable the ICNN core network to remain compact, further reducing the computational complexity of solving the optimization problem. We evaluate the performance of our proposed framework by controlling a battery in the Belgian imbalance settlement mechanism using data from 2023 (\Cref{sec:results,sec:conclusion}).

In summary, our main contributions threefold:
\begin{itemize}
    \item We propose an input-convex, data-driven balancing market model to capture uncertainties, nonlinearities, and partially observable effects, thereby improving on the quarter-hourly balancing market clearing model used in~\cite{smets2023strategic,wessel2024risk,madahi2025model} (\Cref{sec:market model});
    \item We integrate the proposed ICNN with the implicit balancing problem and formulate the resulting price-maker optimization problem as an MIQP to account for the influence of battery actions on imbalance prices, advancing beyond the state-of-the-art data-driven imbalance price prediction models introduced in~\cite{bottieau2024logic,lucas2020price,plakas2025prediction,smets2025value} (\Cref{sec:problem formulation,sec:solve problem});
    \item We incorporate trainable preprocessing steps into the proposed ICNN to improve its performance and keep the core network lightweight, which reduces the computational burden of solving the optimization problem (\Cref{sec:market model}).
\end{itemize}

\section{Implicit Balancing Problem Formulation}
\label{sec:problem formulation}
We aim to control batteries in the imbalance settlement mechanism to minimize their imbalance cost by taking strategic out-of-balance positions. This implies buying energy when the imbalance prices are low and selling it when they are high. For this purpose, we cast the implicit balancing problem as an optimization problem given by~\cref{1,2,3,4,5,6,7}. 

\begin{equation}
   \min_{p_t^\text{cha},p_t^\text{dis}} \>\> \> \sum_{t \in \text{LH}} \lambda_{t} \left( p_t^\text{cha} - p_t^\text{dis} \right) \Delta t
    \label{1}
\end{equation}
\vspace{-4ex}
\begin{align}
    &\text{Subject to:} \notag \\
    &\textrm{SOC}_{t+1} = \textrm{SOC}_t + \left(p_t^\text{cha} \> \eta_\text{cha} - \frac{p_t^\text{dis}}{\eta_\text{dis}} \right) \frac{\Delta t}{E_\text{b}} && \forall t \in \text{LH} \label{2} \\
    &\underline{\textrm{SOC}} \leq \textrm{SOC}_t \leq \overline{\textrm{SOC}} && \forall t \in \text{LH} \label{3} \\
    &0 \leq p_t^\text{cha} \leq z_t^\text{BESS} \> P_b && \forall t \in \text{LH} \label{4} \\
    &0 \leq p_t^\text{dis} \leq \left(1-z_t^\text{BESS}\right) \> P_b && \forall t \in \text{LH} \label{5} \\
    &z_t^\text{BESS} \in \{0,1\} && \forall t \in \text{LH} \label{6} \\ 
    &\lambda_t = f_\text{market}(\myadd{\lvert p_t^\text{cha}-p_t^\text{dis} \rvert, \operatorname{sgn}(p_t^\text{cha}-p_t^\text{dis}),\myadd{\vec{\xi_t}}}) && \forall t \in \text{LH}\label{7}
\end{align}
The objective function~\cref{1} minimizes the imbalance cost over the defined look-ahead horizon (LH). In~\cref{1}, $p_t^\text{cha}$ and $p_t^\text{dis}$ are optimization variables that indicate the charge and discharge power of the battery at time $t$. \Cref{2} defines the battery state of charge (SoC), where $\eta_\text{cha}$ and $\eta_\text{dis}$ denote charging and discharging efficiencies, $E_b$ represents the maximum capacity of the battery, and $\text{SOC}_t$ the battery SoC at time $t$. \Cref{3} ensures that the battery SoC remains within its limits, where $\underline{\text{SOC}}$ and $\overline{\text{SOC}}$ are the minimum and maximum SoC of the battery. \Cref{4,5,6} determine battery's power limits and prevent simultaneous charging and discharging. $P_b$ denotes the maximum power of the battery. \Cref{7} is used to calculate the imbalance price at time $t$ ($\lambda_t$), considering the effect of the battery action. In~\cref{7}, $f_\text{market}$ indicates the balancing market model, \myadd{$\lvert p_t^\text{cha}-p_t^\text{dis} \rvert$ and $\operatorname{sgn}(p_t^\text{cha}-p_t^\text{dis})$ are absolute and sign of the battery action, and $\vec{\xi_t}$} represents inputs that affect price formation (such as the system imbalance) that are independent of the decisions of the BESS operator.

The problem formulation above is not convex as a result of the objective function and~\cref{7}. Using proposition 1 below, we reformulate the objective function in a convex form.

\textbf{Proposition 1.} The function $xf(x,\vec{y})$ is convex in $x$ for $x \ge 0$, provided that $f$ is convex and monotonically increasing with respect to $x$.
\begin{proof}
See Appendix A.
\end{proof}

Using Proposition 1, we rewrite the optimization problem as follows:

\begin{equation}
   \min_{p_t^\text{cha},p_t^\text{dis}} \>\> \> \sum_{t \in \text{LH}}  \left(p_t^\text{cha}\lambda_{t} + p_t^\text{dis}(\underbrace{-\lambda_{t}}_{\lambda'_{t}}) \right) \Delta t
    \label{1b}
\end{equation}
\vspace{-4ex}
\begin{align}
    \hspace{-1cm}\text{Subject to:} & \notag \\
    &\text{\hspace{2cm}  \cref{2,3,4,5,6}} \notag \\
    &\lambda_t = f_\text{market}(p_t^\text{cha},1,\myadd{\vec{\xi_t}}) && \forall t \in \text{LH} \label{7b} \\
    &\lambda'_{t} = f_\text{market}(p_t^\text{dis},-1,\myadd{\vec{\xi_t}}) && \forall t \in \text{LH} \label{7c}
\end{align}

Since charging and discharging do not occur simultaneously, the market model can be decomposed into two independent functions to determine the impact of charging and discharging on the imbalance price. In this paper, we employ a single balancing market model ($f_\text{market}$) for both charging and discharging, with a flag indicating the battery mode (where 1 represents charging). \myadd{The resulting market prices will subsequently be used to determine charging and discharging actions.} By formulating the market model as a convex and monotonically increasing function (see~\cref{subsec: proposed ICNN} for details), the \myadd{objective function of the} reformulated implicit balancing problem becomes convex. Consequently, the global optimum of the formulated MIQP is guaranteed to be found using the branch-and-bound algorithm~\cite{fletcher1998numerical}.

\section{Balancing Energy Market Model}
\label{sec:market model}
\subsection{European Imbalance Settlement Mechanism}
TSOs are primarily responsible for ensuring that the grid frequency remains within the desired range. To achieve this, they correct system imbalances in real-time through the activation of FRR volumes. At the end of each ISP (usually 15\,min), BRPs are settled at imbalance prices that reflect the cost of these activated FRR volumes~\cite{pavirani2025predicting}. As stated in the EBGL, the imbalance settlement mechanism aims to encourage BRPs to restore grid balance by adopting a single pricing methodology~\cite{ENTSO}. This pricing mechanism exposes all BRPs (whether in shortage or in surplus) to an identical price. In this way, the imbalance settlement incentivizes BRPs that help reduce the system imbalance while imposing penalties on those that exacerbate it. The EBGL permits BRPs to deliberately deviate from their nominations to gain an imbalance profit. Implicit balancing benefits BRPs and is also promoted by some TSOs, since it helps reduce system imbalance and lower FRR activation costs. We chose the Belgian imbalance settlement mechanism as a case study in this paper because it is largely consistent with the target model specified in the EBGL. The Belgian imbalance price calculation is explained in detail in~\cite{madahi2025gaming,elia2023brp}.

\subsection{Proposed Data-driven Balancing Market Model}
\label{subsec: proposed ICNN}

\begin{figure}[t]
    \centering
    \includegraphics[width=\linewidth]{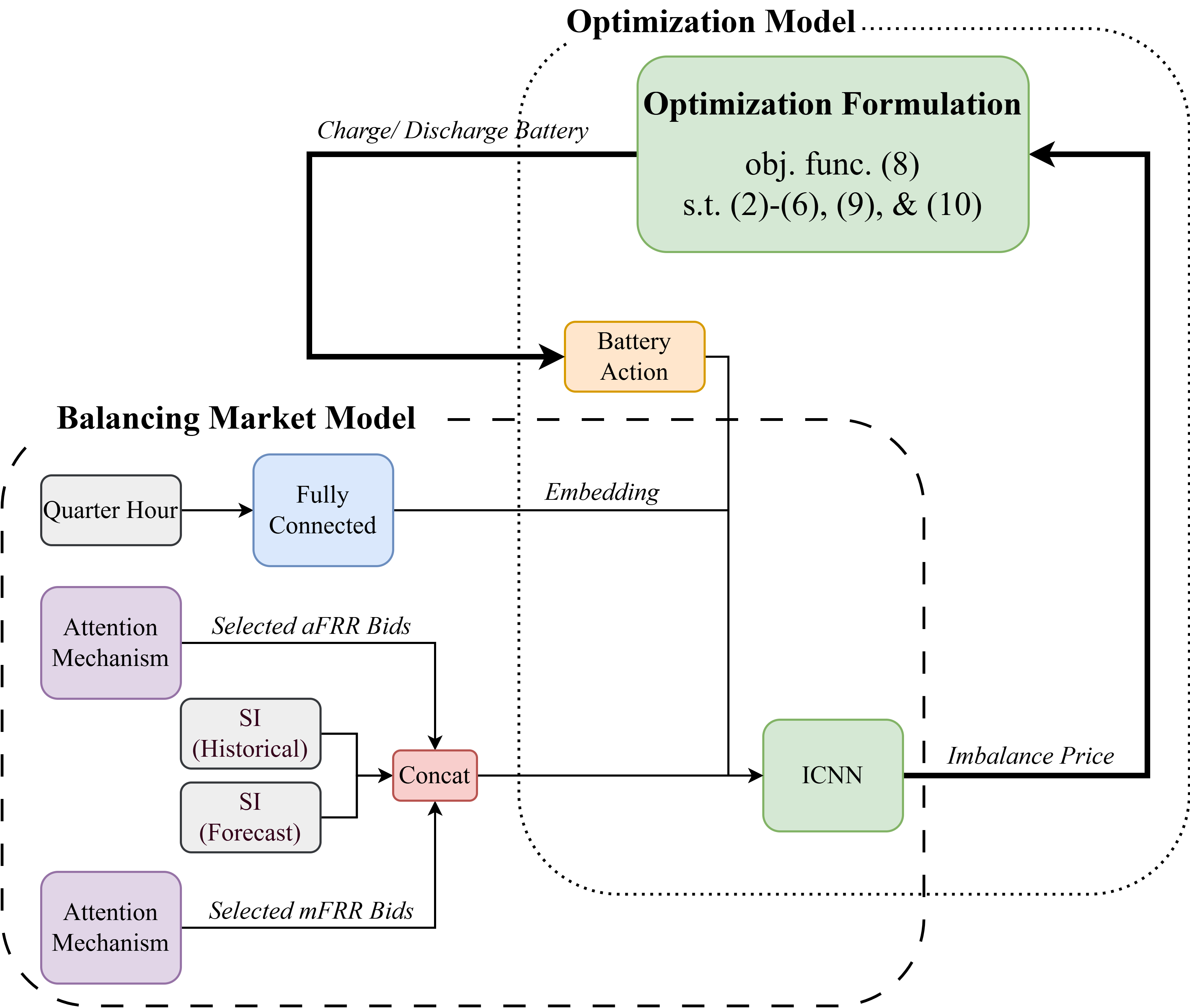}
    \caption{The proposed implicit balancing framework.}
    \label{fig:framework}
\end{figure}

An overview of the proposed framework is presented in \cref{fig:framework}. The proposed model estimates the imbalance price for each quarter hour based on the forecasted system imbalance (SI) for that period, the battery action (determined by solving the optimization problem), and other inputs shown in~\cref{fig:framework}. Historical system imbalances (in this paper, the last two quarter hours) are provided as inputs to the model to better capture market dynamics and the strategies of other market participants. The proposed model consists of trainable preprocessing layers (i.e., embedding and attention layers) and an ICNN. The main purpose of the preprocessing layers is to encode key information and filter out irrelevant data. Following Van Gompel et al.~\cite{van2025probabilistic}, the time feature (quarter hour) is fed into an embedding layer to effectively extract useful information. 

\begin{figure}[t]
    \centering
    \includegraphics[width=\linewidth]{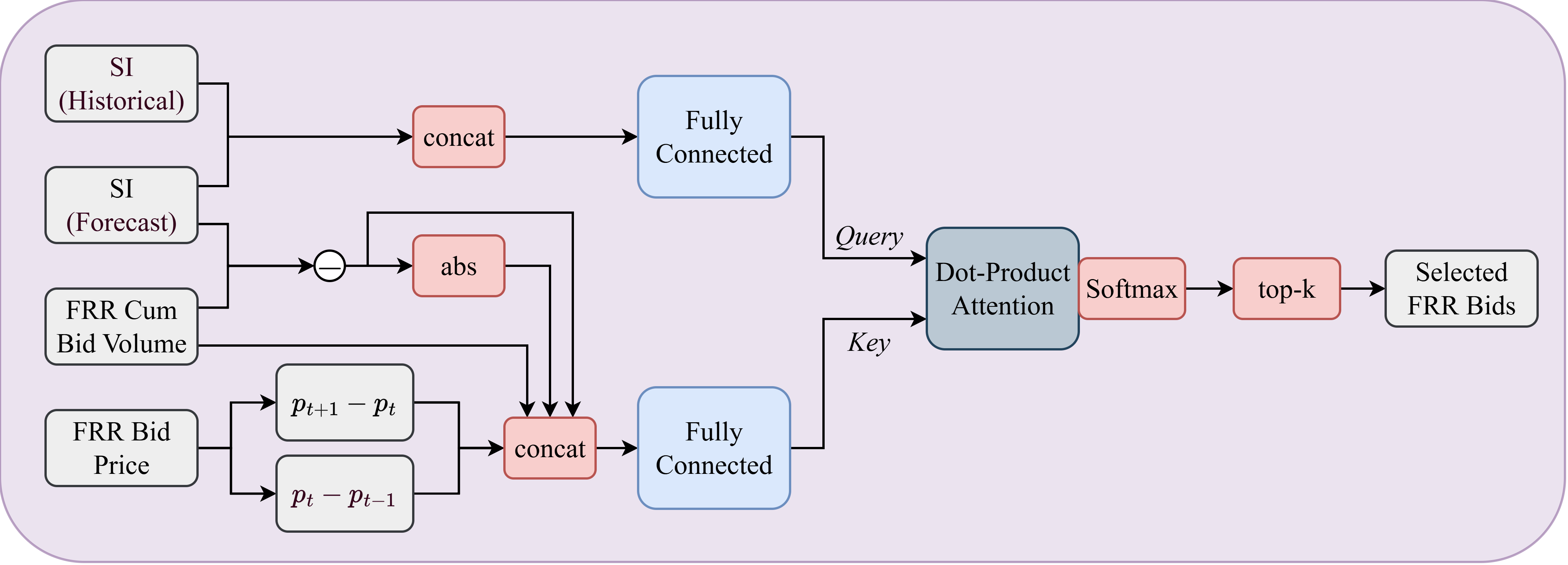}
    \caption{The attention layer architecture used in the proposed market model.}
    \label{fig:attention arch}
\end{figure}

\Cref{fig:attention arch} illustrates the attention block used in the proposed model. In this attention layer, the query vector includes SI-related data (forecasted and historical system imbalances), while the key vectors contain bid-related data (price differences between each bid and its neighboring bids, the signed and absolute distance between bid cumulative volumes and the forecasted SI, and the cumulative bid volumes). The underlying idea is that the network learns the importance of input bids based on the system imbalance inputs. The importance weights for all bids are computed by calculating the dot product of the query vector with all key vectors, followed by the softmax function. Finally, the $k$ bids (64 bids for automatic frequency restoration reserve (aFRR) and 8 bids for mFRR) with the highest weights are selected to be fed into the ICNN. To make the top-k operation differentiable and enable backpropagation through it, we apply the straight-through estimator trick~\cite{bengio2013estimating}. 

\begin{figure}[t]
    \centering
    \includegraphics[width=0.54\linewidth]{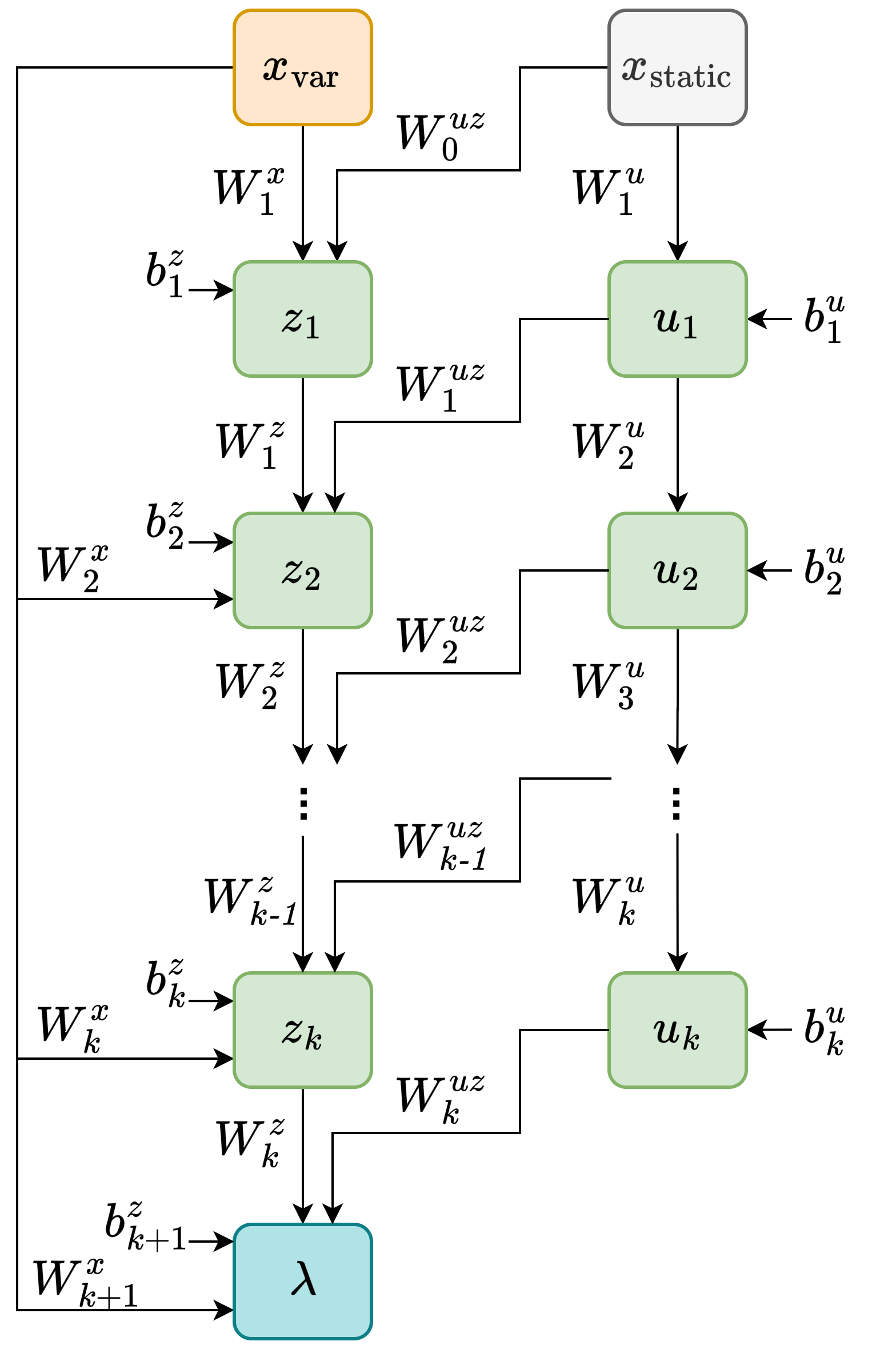}
    \caption{The ICNN architecture.}
    \label{fig:icnn arch}
\end{figure}

The ICNN, shown in~\cref{fig:icnn arch}, is the core component of the proposed model. An ICNN is a neural network in which its weights are constrained such that the network is convex in (a subset of) its inputs~\cite{amos2017input}. We use a partial ICNN, as we require the model to be convex only with respect to certain inputs (i.e., the battery's charge and discharge actions). The ICNN defined in~\cref{fig:icnn arch} maps the inputs ($x_\text{var}$ and $x_\text{static}$) to the outputs ($\lambda$ or the imbalance price) as follows:
\begin{align}
    & z_i=\sigma_i(W^z_{i-1} z_{i-1}+W^x_i x_\text{var}+W^{uz}_{i-1} u_{i-1}+b^z_i) \label{11} \\
    & u_i=\tilde{\sigma}_i(W^u_i u_{i-1}+b^u_i) \\
    & u_0 = x_{\text{static}}, \,  z_{k+1} = \lambda, \, W^z_0 = \mathbf{0}
\end{align}
where $z_i$ and $u_i$ are hidden units, $W$ and $b$ are trainable weights and biases, $\sigma_i$ and $\tilde{\sigma_i}$ are activation functions (ReLU in this paper, except for $\sigma_{k+1}$ which is linear), $x_\text{var}$ denotes the inputs with respect to which the network is convex, and $x_\text{static}$ represents the remaining inputs. The following propositions ensure that the defined ICNN is convex and increases monotonically with respect to $x_\text{var}$.

\textbf{Proposition 2.} The network $f(x_\text{var},x_\text{static})$ is convex in $x_\text{var}$ if all elements in $W^z_i$ ($\forall i \leq k$) are nonnegative and $\sigma_i$ ($\forall i \leq k+1$) is convex and nondecreasing.
\begin{proof}
As shown in~\cite{amos2017input}.
\end{proof}

\textbf{Proposition 3.} The network $f(x_\text{var},x_\text{static})$ is monotonically increasing with respect to $x_\text{var}$ if all elements in $W^x_i$ ($\forall i \leq k+1$) are nonnegative and $\sigma_i$ ($\forall i \leq k+1$) is nondecreasing.
\begin{proof}
See Appendix B.
\end{proof}

The assumption that the output of the proposed market model (the imbalance price) monotonically increases with respect to battery actions aligns with what occurs in practice: charging the battery causes a negative imbalance, which increases the upward FRR volume activation (or reduces the downward FRR volume activation), resulting in an increase in the imbalance price. On the other hand, as the discharge power increases, the imbalance price decreases accordingly. In~\cref{7c}, since our proposed model outputs the negative of the price, it remains consistent with this monotonically increasing relationship between discharge power and output of the market model (i.e., the negative of the imbalance price). 

\section{Solve The Implicit Balancing Problem}
\label{sec:solve problem}
We employ MPC, a receding horizon method, to solve the implicit balancing problem formulated in~\cref{1b,2,3,4,5,6,7b,7c}. The proposed ICNN can be formulated as a mixed-integer linear program (MILP) by modeling ReLU activation functions using binary variables~\cite{fischetti2018deep}. The other parts of the proposed market model (the purple and blue blocks in~\cref{fig:framework}) do not need to be converted into an MILP, as their parameters and inputs are constant with respect to the decision variables ($p_t^\text{cha}$ and $p_t^\text{dis}$).

We use Gurobi to obtain a solution to the resulting MIQP. \myadd{The overall problem is nonconvex due to binary variables ($z_t^\text{BESS}$). However, since the number of binary variables is relatively small (equal to the look-ahead horizon length and the binaries obtained from the reformulation of the ICNN) divide-and-conquer methods such as the branch-and-bound algorithm can guarantee convergence to the optimal solution, given that each resulting subproblem is convex~\cite{fletcher1998numerical}.} Since the Belgian ISP is set to 15 minutes, the time resolution of decision-making is also 15 minutes. At the beginning of each quarter hour, the optimization problem is solved over the look-ahead horizon, and the resulting battery action for that quarter hour is applied. At each time step, the market model output corresponding to the opposite direction of the taken battery action does not affect the objective function value in~\cref{1b}, as the decision variable associated with the opposite direction is zero (to prevent simultaneous charging and discharging).

\section{Results and Discussion}
\label{sec:results}
\subsection{\myadd{Simulation} Setup}

\begin{table*}[t]
\centering
\caption{Imbalance profit for 1\,MW/\,2\,MWh battery on December 2023.}
\begin{threeparttable}
\begin{tabular}{cccccc}
\midrule \midrule
\multirow{3}{*}{\textbf{Horizon}} & \multirow{3}{*}{\textbf{SI} \textbf{Forecast}} & \multirow{3}{*}{\textbf{Method}} &  \multicolumn{3}{c}{\textbf{Profit (\texteuro/\,MW/\,QH)}} \\ 
\cmidrule(lr){4-6}
{} & {} & {} & {\textbf{All}} & \textbf{$|\text{SI}|\ge$20\,MW} & \textbf{$|\text{SI}|<$20\,MW}\\

\midrule \midrule
\multirow{8}{*}{1 QH} & \multirow{5}{*}{PF} & Optimal & 9.59 & 11.23 & 8.06 \\
\cmidrule(lr){3-6}
 &  & Clearing Approx. & 8.01 & 9.89 & 6.27 \\
 \cmidrule(lr){3-6}
  &  & Proposed & 8.4$\pm$0.08\tnote{*}  & 10.43$\pm$0.15 & 6.49$\pm$0.14 \\
   \cmidrule(lr){2-6}
 & \multirow{2}{*}{Noise} & Clearing Approx. & 5.58 & 10.51 & 1.02 \\
 \cmidrule(lr){3-6}
  &  & Proposed & 7.68$\pm$0.19 & 10.56$\pm$0.15 & 4.99$\pm$0.45 \\
  \cmidrule(lr){1-6}
\multirow{8}{*}{4 QH} & \multirow{5}{*}{PF} & Optimal & 10.84 & 14.02 & 7.9 \\
\cmidrule(lr){3-6}
 &  & Clearing Approx. & 9.7 & 12.58 & 7.03 \\
 \cmidrule(lr){3-6}
  &  & Proposed & 9.59$\pm$0.14 & 13.31$\pm$0.35 & 6.14$\pm$0.13 \\
   \cmidrule(lr){2-6}
 & \multirow{2}{*}{Noise} & Clearing Approx. & 6.55 & 12.87 & 0.64 \\
 \cmidrule(lr){3-6}
  &  & Proposed & 8.87$\pm$0.23 & 13.46$\pm$0.4 & 4.60$\pm$0.44 \\
\midrule \midrule
\end{tabular}
\begin{tablenotes}
    \item[*] Standard deviation over the 10 random seeds
\end{tablenotes}
\end{threeparttable}
\label{table:1MW results}
\end{table*}

We assess the performance of the proposed NN-assisted MPC in terms of price prediction and imbalance profit using the 2023 Belgian balancing dataset. We study both the price-taker and price-maker scenarios: the former to quantify the gap between the obtained and optimal imbalance profits, and the latter to investigate the impacts of battery actions on price prediction error and imbalance profit. The dataset is divided as follows: the first 10 months are used for training, November for validation, and December for testing. Four different battery configurations are studied to analyze the effect of battery size: 1\,MW/\,2\,MWh, 10\,MW/\,20\,MWh, 50\,MW/\,100\,MWh, and 100\,MW/\,200\,MWh, all with a 90\% round-trip efficiency for both charging and discharging. The proposed ICNN consists of two hidden layers with 64 and 32 neurons for the network using $x_\text{var}$ as inputs, and two hidden layers with 1024 and 512 neurons for the network using $x_\text{static}$ as inputs. The quarter-hour feature shown in~\cref{fig:framework} is embedded into a vector of size 4 using a fully connected network with two hidden layers of 256 and 128 neurons. The proposed data-driven balancing market model is trained end-to-end using the Adam optimizer with an L1 loss function and a learning rate of \num{2e-4} for 40 epochs. The proposed model is trained with 10 random seeds, and the reported results are averaged over all seeds to ensure reliability. All experiments were executed on a 6-core Intel Core i5 (2.90 GHz) machine with 32 GB of RAM.

For both upward and downward directions, aFRR and mFRR bids are separately aggregated to form the available regulation capacity (ARC) merit order for each quarter hour. The resulting ARC merit orders for aFRR and mFRR are discretized in 2\,MW and 100\,MW steps. In most cases, we assume perfect foresight (PF) on the system imbalance to isolate the impact of the balancing market models on MPC decision quality. To study the robustness of the proposed framework to imperfect system imbalance forecasts, we also show results when introducing zero-mean Gaussian noise with a standard deviation of 10\,MW, increasing exponentially by 20\% across the optimization look-ahead horizon. 

The proposed model is benchmarked against the quarter-hour-level, optimization-based balancing market clearing model (hereafter referred to as the clearing approximation model in this paper), as detailed in~\cite{smets2023strategic,madahi2025model}. The clearing approximation model is widely used due to its simplicity and convexity. It estimates the imbalance price using the average system imbalance over the quarter hour, neglecting sub-quarter-hourly dynamics and changes in system imbalance within that quarter hour. 

The final imbalance profit of the battery, resulting from the MPC actions of both the proposed and clearing approximation models, is calculated by implementing these actions on a real-minute-based balancing market simulator~\cite{elia2023brp}.\footnote{Although the Belgian TSO calculates imbalance prices through a 4-second optimization cycle for aFRR activations, the available balancing data (system imbalance, aFRR/mFRR activations, etc.) are published at one-minute intervals. Thus, our balancing market simulator uses a one-minute resolution as the finest available granularity.} 

\subsection{Price-taker Battery}

\begin{figure}[t]
    \centering
    \includegraphics[width=0.94\linewidth]{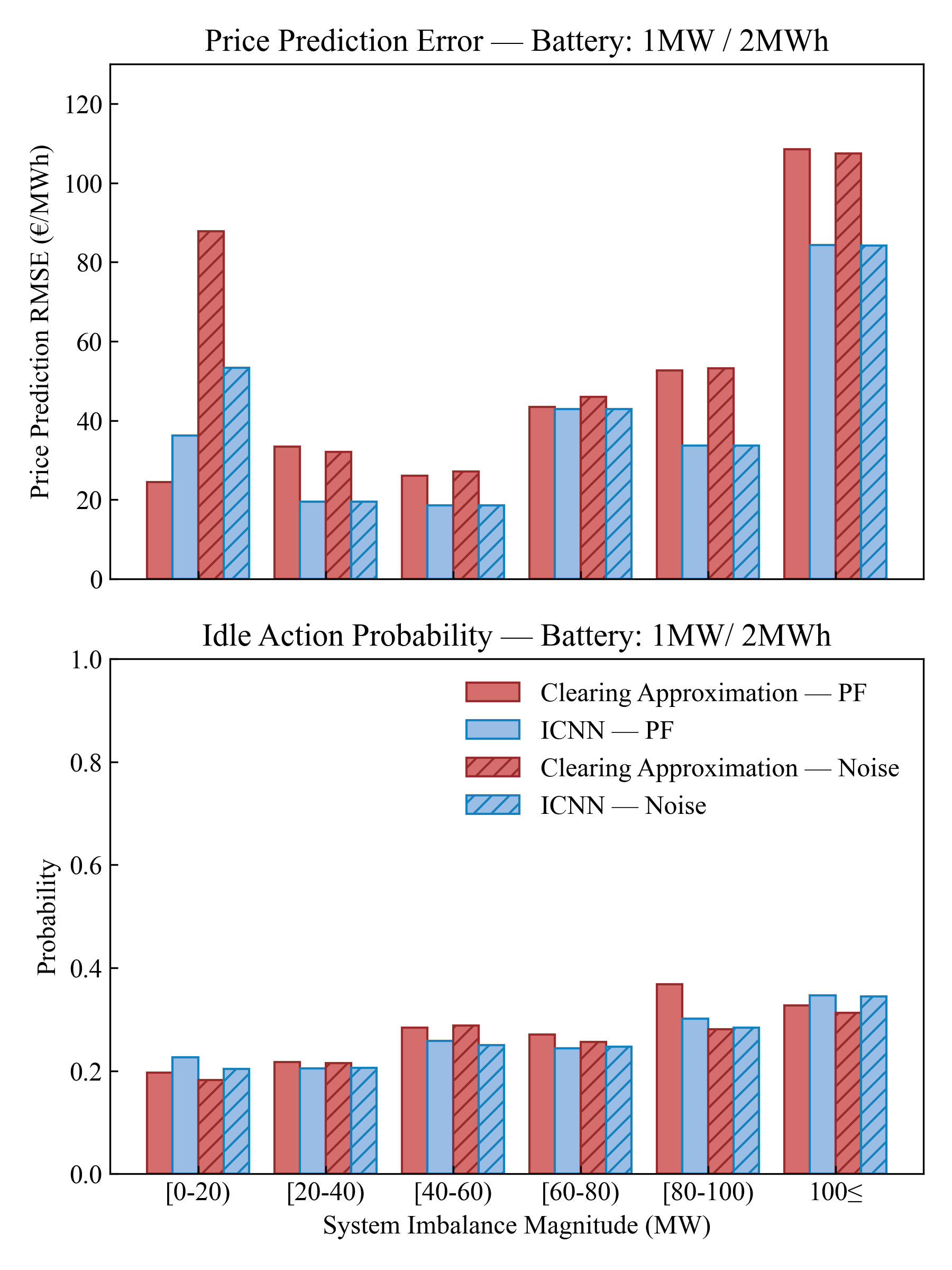}
    \caption{Price prediction error and the probability of idle action for 1\,MW/\,2\,MWh battery.}
    \label{fig:si bin 1MW}
\end{figure}

The proposed MPC framework is applied to a 1\,MW/\,2\,MWh battery. Since the battery is small-scale in this case, it cannot significantly influence imbalance prices and is therefore a price-taker. Consequently, the optimal imbalance profit can be obtained for this case. \Cref{table:1MW results} summarizes the imbalance profits for different methods and look-ahead horizons. The results show that when both the clearing approximation and the proposed models have access to the exact SI values over their horizon, their MPC performance is close to optimal (i.e., approximately 11\% lower than the optimal profit for a 4-quarter-hour (QH) horizon). As~\cref{fig:si bin 1MW} demonstrates, the proposed model improves the average price prediction RMSE for quarter hours with large SI ($|\text{SI}| \ge 20$\,MW) by 23\% compared to the clearing approximation model, whereas for quarter hours with small SI ($|\text{SI}|<20$\,MW), the benchmark model performs better in terms of RMSE. This aligns with the results shown in~\cref{table:1MW results}, where the proposed method outperforms the clearing approximation model for large SI, while the latter makes higher profit for quarter hours with small SI. 

Adding noise to SI values significantly reduces the performance of the clearing approximation model, particularly for small SI values. This occurs because adding noise to SI values in this range can reverse the direction of the SI. Consequently, the noisy values mislead the benchmark model about the correct direction of the SI, leading to a large prediction error. On the other hand, the proposed method is robust to noise, with its performance reduced by about 8\% compared to the perfect foresight case. In this case, the proposed method outperforms the clearing approximation model by 35.4\% in terms of imbalance revenue when the look-ahead horizon is four quarter-hours.

\subsection{Price-maker Battery}

\begin{table*}[t]
\centering
\caption{Imbalance profit for large batteries with 4 QH look-ahead horizon on December 2023.}
\begin{threeparttable}
\begin{tabular}{cccccc}
\midrule \midrule
\multirow{3}{*}{\textbf{Battery}} & \multirow{3}{*}{\textbf{SI} \textbf{Forecast}} & \multirow{3}{*}{\textbf{Method}} &  \multicolumn{3}{c}{\textbf{Profit (\texteuro/\,MW/\,QH)}} \\ 
\cmidrule(lr){4-6}
{} & {} & {} & {\textbf{All}} & \textbf{$|\text{SI}|\ge$20\,MW} & \textbf{$|\text{SI}|<$20\,MW}\\

\midrule \midrule
 \multirow{6}{*}{10\,MW/ 20\,MWh} & \multirow{3}{*}{PF} & Clearing Approx. & 6.41 & 11.04 & 2.59 \\
 \cmidrule(lr){3-6}
  &  & Proposed & 6.93$\pm$0.22\tnote{*} & 11.73$\pm$0.37 & 2.98$\pm$0.27 \\
   \cmidrule(lr){2-6}
 & \multirow{2}{*}{Noise} & Clearing Approx. & 3.54 & 10.51 & $-$2.36 \\
 \cmidrule(lr){3-6}
  &  & Proposed & 6.83$\pm$0.18 & 11.72$\pm$0.36 & 2.79$\pm$0.22 \\
  \cmidrule(lr){1-6}

\multirow{6}{*}{50\,MW/ 100\,MWh} & \multirow{3}{*}{PF} & Clearing Approx. & 2.76 & 6.19 & 0.49 \\
 \cmidrule(lr){3-6}
  &  & Proposed & 3.6$\pm$0.13 & 6.86$\pm$0.23 & 1.28$\pm$0.14 \\
   \cmidrule(lr){2-6}
 & \multirow{2}{*}{Noise} & Clearing Approx. & 1.32 & 5.30 & $-$1.67 \\
 \cmidrule(lr){3-6}
  &  & Proposed & 3.59$\pm$0.13 & 6.88$\pm$0.24 & 1.26$\pm$0.13 \\
  \cmidrule(lr){1-6}

\multirow{6}{*}{100\,MW/ 200\,MWh} & \multirow{3}{*}{PF} & Clearing Approx. & 1.36 & 3.79 & 0.04 \\
 \cmidrule(lr){3-6}
  &  & Proposed & 2.21$\pm$0.04 & 4.15$\pm$0.12 & 0.42$\pm$0.07 \\
   \cmidrule(lr){2-6}
 & \multirow{2}{*}{Noise} & Clearing Approx. & 0.55 & 3.30 & $-$1.19 \\
 \cmidrule(lr){3-6}
  &  & Proposed & 2.21$\pm$0.04 & 4.15$\pm$0.12 & 0.43$\pm$0.1 \\

\midrule \midrule
\end{tabular}
\begin{tablenotes}
    \item[*] Standard deviation over the 10 random seeds
\end{tablenotes}
\end{threeparttable}
\label{table:large batteries}
\end{table*}

\begin{figure}[t]
    \centering
    \includegraphics[width=0.94\linewidth]{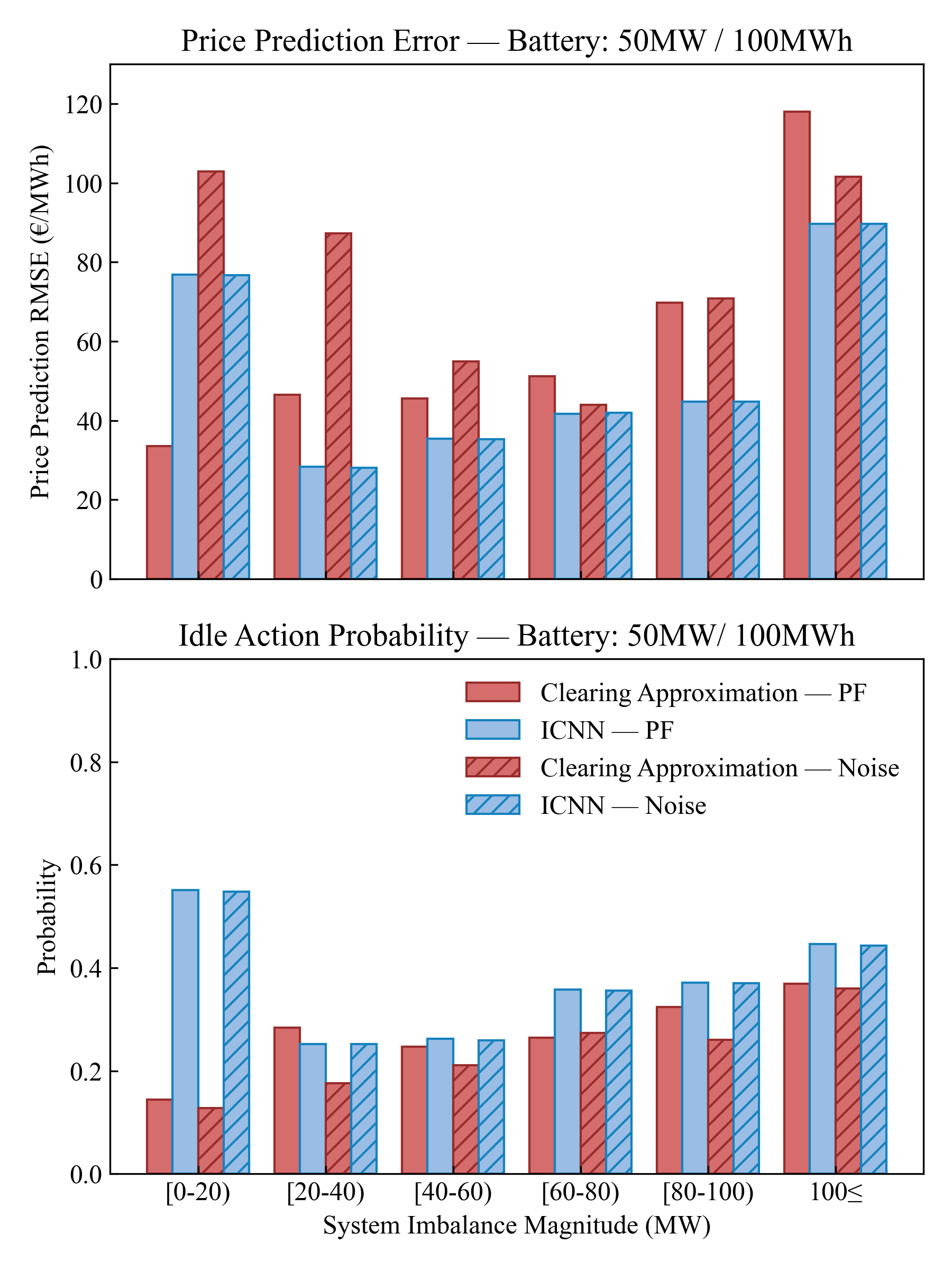}
    \caption{Price prediction error and the probability of idle action for 50\,MW/\,100\,MWh battery.}
    \label{fig:si bin 50MW}
\end{figure}

\begin{figure}[t]
    \centering
    \begin{subfigure}{0.48\textwidth}
        \includegraphics[width=\linewidth]{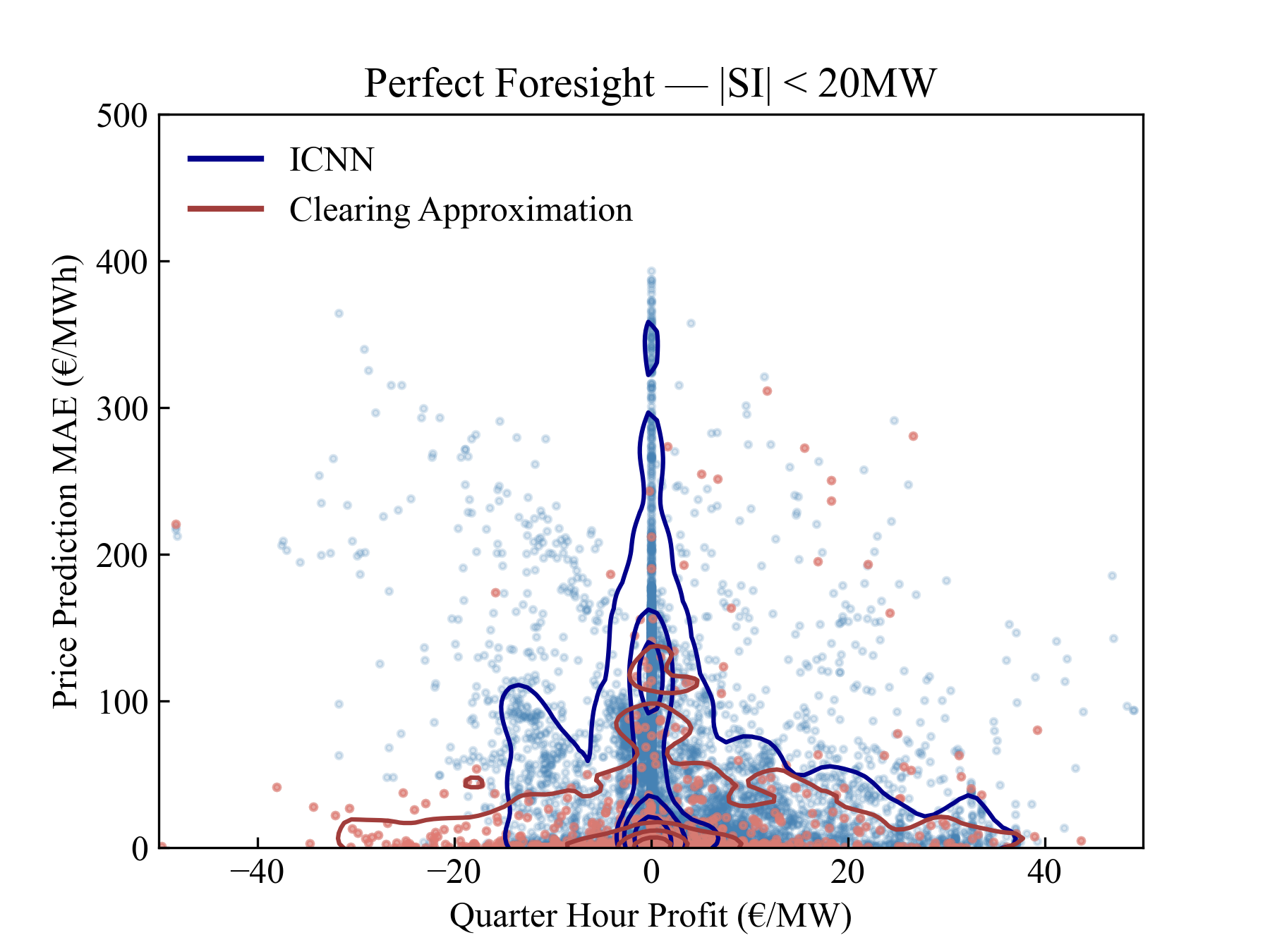}
        \caption{}
        \label{fig:profit spread pf}
    \end{subfigure}
    \begin{subfigure}{0.48\textwidth}
        \includegraphics[width=\linewidth]{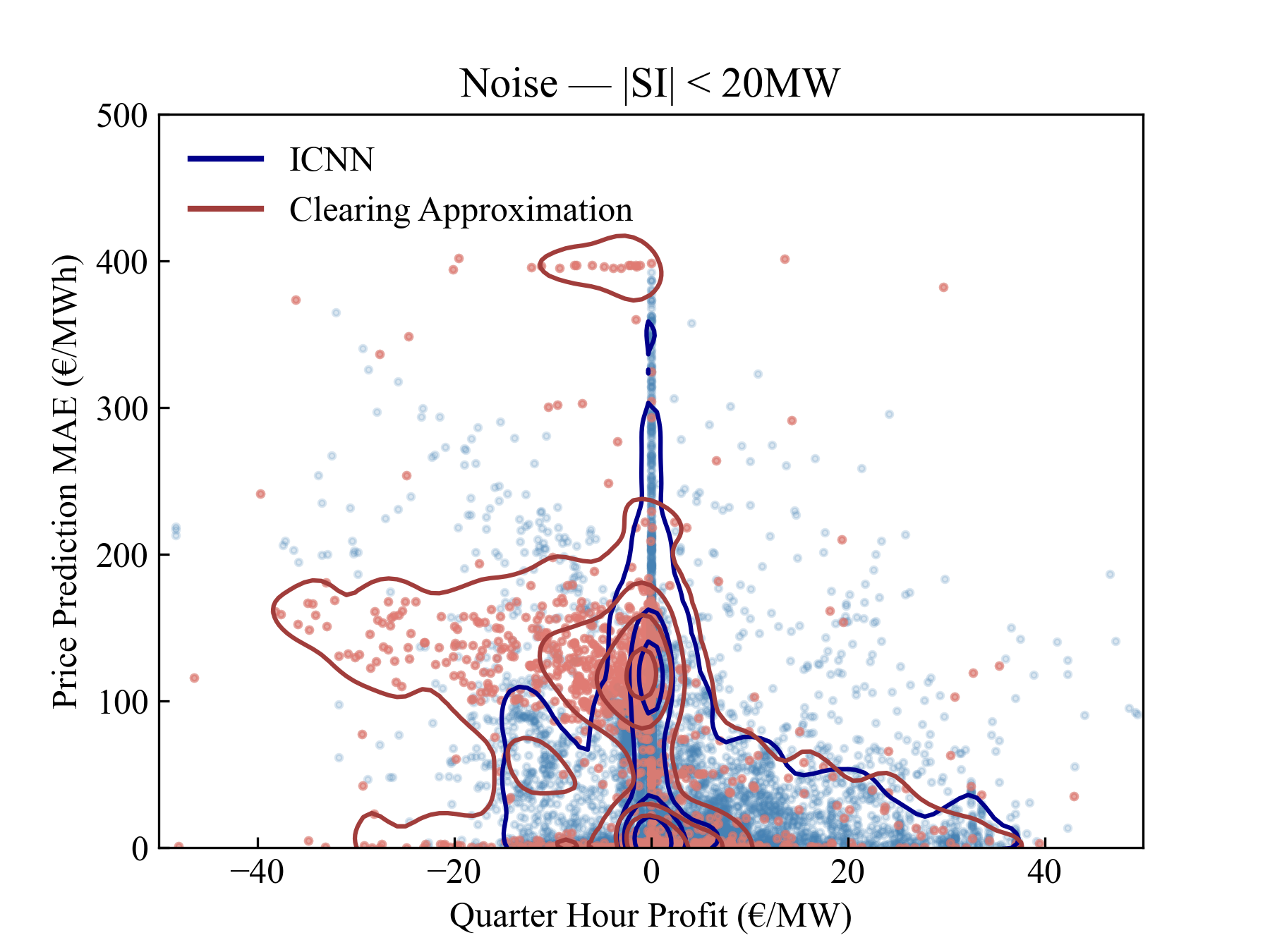}
        \caption{}
        \label{fig:profit spread noise}
    \end{subfigure}
    \caption{The QH profit spread of the 50\,MW battery for quarter hours with small SI (a) PF (b) noise.}
    \label{fig:profit spread}
\end{figure}

For large-scale batteries, whose impact on the grid is greater, sub-quarter-hour variations in SI become increasingly important in determining the final imbalance price. This leads to an increase of 11\% in the price prediction error of the clearing approximation model for the 50\,MW battery compared to the price-taker case, as illustrated in~\cref{fig:si bin 50MW}. For quarter hours with large SI values, the prediction error of the clearing approximation model increases by 15.5\% compared to the price-taker case, whereas for the proposed method, this deterioration is limited to 6.8\%. In contrast, the prediction error of the proposed model increases substantially for quarter hours with small SI. However, the proposed model yields a higher imbalance profit for these quarter hours compared to the benchmark model, as indicated in~\cref{table:large batteries}.  The reason is that the proposed framework executes noticeably more idle actions during these quarter hours. \Cref{fig:profit spread} shows that the quarter hour profits for quarter hours with small SI in the proposed model are mostly centered close to zero, in spite of their high prediction error. The positive tails of the contours for both models are almost the same, which indicates that the quarter hours in which the proposed method takes an idle action are those in which the benchmark model takes an incorrect non-idle action, leading to negative profits. For quarter hours with large SI, the proposed method achieves superior imbalance profits than the benchmark, owing to its lower prediction error. In general, the proposed method improves the imbalance profit by 8.1\%, 30.4\%, and 62.5\% for 10\,MW, 50\,MW, and 100\,MW batteries, respectively, compared to the benchmark.

The results in~\cref{table:large batteries} confirm that, similar to the price-taker case, our proposed framework is robust to noise. Also, as the battery size increases, the proposed method becomes less sensitive to noise because it takes more idle actions during quarter hours with small SI values: for these quarter hours, the proportion of idle actions increases from 0.23 in the price-taker case to 0.55 for the 50 MW battery. 

Finally, Replacing the clearing approximation model with the proposed model can speed up the optimization process, regardless of battery size. The computational time per quarter hour is reduced on average from 60.59\,ms in the benchmark model to 30.3\,ms in the proposed model. 






\section{Conclusion}
\label{sec:conclusion}
We propose an NN-assisted MPC framework for controlling batteries in the imbalance settlement mechanism. We model the balancing energy market using a data-driven ICNN-based network to efficiently capture uncertainties and partial observability, ensure the convexity of \myadd{the objective function of} the resulting optimization problem, and achieve low computational time. We compare our proposed framework against the commonly used quarter-hour-level market clearing model approximation, in a case study using 2023 Belgian balancing data. Our proposed framework using the ICNN outperforms
an MPC that uses the baseline market clearing model to make decisions, achieving 8.1\% to 62.5\% higher imbalance profits across different battery sizes. Although the benchmark model has smaller price prediction errors for quarter hours with small system imbalance values (below 20\,MW), our proposed framework still achieves higher overall imbalance profits for these quarter hours, as it more often resorts to the idle action\,---\,particularly in the price-maker case, when a large battery system is used. For both price-taker and price-maker scenarios, the imbalance profit of our proposed framework is less sensitive to noise added to the system imbalance forecasts. In this case, the proposed framework achieves 35.4\% to 301.8\% higher imbalance revenue for different battery sizes relative to the clearing approximation model. Overall, our proposed data-driven market model helps the MPC make better decisions while reducing computational time by 50\%.

The current work can be extended by enriching the inputs to the proposed model with longer historical data and additional grid-related features, such as renewable forecast errors. A deep neural network can be used to extract useful information from these inputs and feed it to the ICNN, which could potentially improve model accuracy. 
Moreover, since the ICNN is lightweight and capable of modeling nonlinearities, our proposed framework enables formulating the implicit balancing problem on a minute basis, allowing it to take advantage of minute-level balancing data and the most recent information for decision-making. Future work includes exploring such minute-based battery control actions. \myadd{In this work, we use a simple linear battery model. Another direction for future work is to replace this basic battery model with a one that captures the battery behavior more accurately, but still retains convexity.}

\section*{Acknowledgments}
This research was partly funded by the Flemish Government under the ``Onderzoeksprogramma Artifici\"ele Intelligentie (AI) Vlaanderen'' programme as well as a travel grant from the Research Foundation -- Flanders (FWO).

\section*{Appendix A}
\begin{proof}
Let $\alpha \in (0,1)$ and $x_1, \,x_2 \ge 0$. As $f$ is a monotonically increasing function, we can write:
\begin{equation}
    \alpha(1-\alpha)(x_2-x_1)(f(x_2,\vec{y})-f(x_1,\vec{y})) \ge 0
    \label{A1}
\end{equation}

By expanding~\cref{A1}, we obtain the following inequality.
\begin{align}
    (\alpha x_1+(1-\alpha)x_2) & \Big(( \alpha f(x_1,\vec{y})+(1-\alpha)f(x_2,\vec{y}) \Big)\le  \notag \\ 
    &\alpha x_1 f(x_1,\vec{y}) + (1-\alpha) x_2 f(x_2,\vec{y})
    \label{A2}
\end{align}

On the other hand, since $f$ is convex, the definition of convexity gives the following inequality.
\begin{align}
    f(\alpha x_1 + (1-\alpha) x_2,\vec{y})\le \alpha f(x_1,\vec{y}) + (1-\alpha) f(x_2,\vec{y})
    \label{A3}
\end{align}

Multiplying~\cref{A3} by $(\alpha x_1+(1-\alpha)x_2)$ (which is nonnegative) yields the following inequality.
\begin{align}
    (\alpha x_1+(1-\alpha)x_2)&f(\alpha x_1 + (1-\alpha) x_2,\vec{y}) \le  \notag \\ 
    (\alpha x_1+(1-\alpha)x_2) &\Big(( \alpha f(x_1,\vec{y})+(1-\alpha)f(x_2,\vec{y}) \Big)
    \label{A4}
\end{align}

Combining~\cref{A2,A4} results in~\cref{A5}.
\begin{align}
    (\alpha x_1+(1-\alpha)x_2)&f(\alpha x_1 + (1-\alpha) x_2,\vec{y}) \le \notag \\
    \alpha &x_1 f(x_1,\vec{y}) + (1-\alpha) x_2 f(x_2,\vec{y})
    \label{A5}
\end{align}

In \cref{A5}, the convexity of $xf(x,\vec{y})$ is demonstrated.
\end{proof}

\section*{Appendix B}
\begin{proof}
Consider $x_\text{var} \ge x'_\text{var}$. For $i=1$ in~\cref{11}, the following inequality can be written.
\begin{align}
    W^x_1 x_\text{var} + W^{uz}_0 x_\text{static}+b^z_1 \ge W^x_1 x'_\text{var} + W^{uz}_0 x_\text{static}+b^z_1
    \label{B1}
\end{align}

Since $\sigma_1$ is nondecreasing, applying it to both sides of the inequality in~\cref{B1} yields $z_1 \ge z'_1$. Similarly, for $i \ge 2$ in~\cref{11}, we can obtain the following inequality.

\begin{align}
    \underbrace{\sigma_i(W^z_{i-1} z_{i-1}+W^x_i x_\text{var}+W^{uz}_{i-1} u_{i-1}+b^z_i)}_{z_i} \ge \notag \\ 
    \underbrace{\sigma_i(W^z_{i-1} z_{i-1}+W^x_i x'_\text{var}+W^{uz}_{i-1} u_{i-1}+b^z_i)}_{z'_i}
    \label{B2}
\end{align}

Using~\cref{B2}, $z_{k+1}=f(x_\text{var},x_\text{static}) \ge z'_{k+1}=f(x'_\text{var},x_\text{static})$, which demonstrates the monotonicity of $f$.
\end{proof}

\bibliographystyle{IEEEtran}
\bibliography{reference}

\end{document}